\documentclass{article}
\usepackage[T1]{fontenc}
\usepackage[utf8]{inputenc}
\usepackage[]{ismir} 
\usepackage{amsmath,cite,url}
\usepackage{graphicx}
\usepackage{color}
\usepackage{caption} 
\usepackage{graphicx}
\usepackage{algorithm}
\usepackage{algorithmic}
\usepackage{booktabs}
\usepackage{multirow}
\usepackage{makecell}
\usepackage{pifont}
\usepackage{enumitem}
\usepackage{subfigure}
\usepackage{amsfonts}

\title{Instruct-MusicGen: Unlocking Text-to-Music Editing
for Music Language Models via Instruction Tuning}

\multauthor
  {Yixiao Zhang$^1$, Yukara Ikemiya$^2$, Woosung Choi$^2$, Naoki Murata$^2$, Marco A. Martínez-Ramírez$^2$,}
  {{\bf Liwei Lin$^3$, Gus Xia$^3$, Wei-Hsiang Liao$^2$, Yuki Mitsufuji$^2$, Simon Dixon$^1$}\\
  $^1$ C4DM, Queen Mary University of London\\
  $^2$ Sony AI \hspace{1cm}
  $^3$ Music X Lab, MBZUAI\\
  {\tt\small {first.last}@qmul.ac.uk, {first.last}@sony.com, \{gus.xia, ll4270\}@nyu.edu}
  }
\def\authorname{Y. Zhang, Y. Ikemiya, W. Choi, N. Murata, M. A. Martínez-Ramírez, L. Lin, G. Xia, W.-H. Liao, Y. Mitsufuji, and S. Dixon}

\usepackage[bookmarks=false,pdfauthor={\authorname},pdfsubject={\pdfsubject},hidelinks]{hyperref}

\begin{document}

\maketitle

\begin{abstract}

  The task of text-to-music editing, which employs text queries to modify music  (e.g.\ by changing its style or adjusting instrumental components), presents unique challenges and opportunities for AI-assisted music creation. Previous approaches in this domain have been constrained by the necessity to train specific editing models from scratch, which is both resource-intensive and inefficient; other research uses large language models to predict edited music, resulting in imprecise audio reconstruction. In this paper, we introduce \textit{Instruct-MusicGen}, a novel approach that finetunes a pretrained MusicGen model to efficiently follow editing instructions such as adding, removing, or separating stems. Our approach involves a modification of the original MusicGen architecture by incorporating a text fusion module and an audio fusion module, which allow the model to process instruction texts and audio input concurrently and yield the desired edited music. Remarkably, although Instruct-MusicGen only introduces $\sim$8\% new parameters to the original MusicGen model and only trains for 5K steps, it achieves superior performance across all tasks compared to existing baselines. This advancement not only enhances the efficiency of text-to-music editing but also broadens the applicability of music language models in dynamic music production environments. \footnote{Code, model weights and demo are available at: \url{https://github.com/ldzhangyx/instruct-musicgen}.} \footnote{This work was done during Yixiao Zhang's internship at Sony AI.}
  
\end{abstract}  

\section{Introduction}\label{sec:intro}

The rapid advances in text-to-music generation have opened up new possibilities for AI-assisted music creation~\cite{musiclm, musicgen, jen1, audioldm2, musicldm}. This paradigm shift has also sparked a growing interest in developing models that offer greater controllability~\cite{cocomulla, musiccontrolnet, mustango, airgen} and editability~\cite{instructME, musicmagus, M2UGen} over the music generation process. In music production, a stem—a mixed group of tracks often related by instrument type (like drums or lead vocals)—is essential for mixing and mastering because it allows producers to isolate, adjust, and manipulate individual elements of a song. Following the definition in MusicMagus~\cite{musicmagus}, ``text-to-music editing" involves using textual queries to modify various aspects of a music recording, which can be categorised into two main types: intra-stem editing, which focuses on modifying a single stem (e.g., changing the instrument, timbre, or performance style), and inter-stem editing, which involves altering the relationships among stems (e.g., adding, removing, or separating stems). Our work mainly focuses on the problem of inter-stem editing. 

\begin{figure}[tbp]
    \centering
    \hspace*{-1cm}    \includegraphics[width=\linewidth, trim=0.5cm 24cm 7.5cm 0cm]{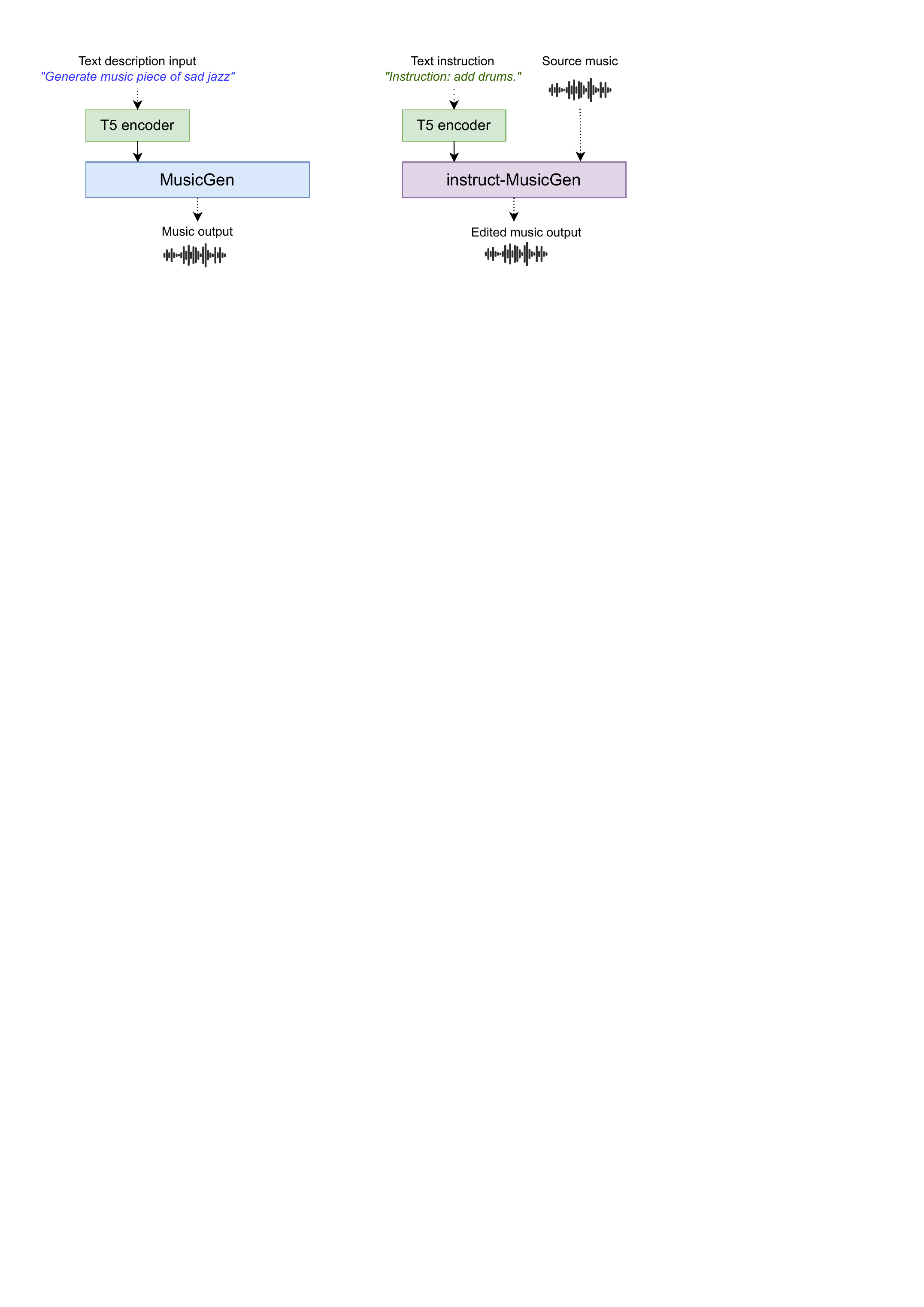}
    \caption{Comparison between MusicGen and instruct-MusicGen. Instruct-MusicGen accepts both audio input and editing instruction text as conditions.}
    \label{fig:preview}
\end{figure}

Previous attempts to develop text-based music editing models have encountered several challenges. Some approaches~\cite{AUDIT, instructME} have focused on training specialised editing models from scratch, which is resource-intensive and may not yield results comparable to state-of-the-art music generation models. Other work~\cite{loopcopilot, M2UGen, uniaudio} has sought to leverage existing large language models (LLMs) and MusicGen~\cite{musicgen}, allowing the LLM to interpret editing instructions without further training the music model. Although this approach offers flexibility, it often lacks the ability to precisely reconstruct the conditional audio, leading to unreliable results. To address these limitations, an ideal solution should harness the knowledge embedded in pretrained models to ensure high-quality audio output while adapting the architecture to accommodate the specific requirements of music editing tasks.

In this paper, we introduce Instruct-MusicGen, a novel approach that applies an instruction-following tuning strategy to the pretrained MusicGen model, enhancing its ability to follow editing instructions effectively without finetuning all its parameters. As shown in Figure~\ref{fig:preview}, by incorporating an audio fusion module based on LLaMA-Adapter~\cite{llamaadapter, cocomulla} and a text fusion module based on LoRA~\cite{lora} into the original MusicGen architecture, we allow the model to process both precise audio conditions and text-based instructions simultaneously, which the original MusicGen does not do. This enables Instruct-MusicGen to perform a range of editing tasks. In this paper, we focus on a specific set of these tasks: adding, separating, and removing stems. To train Instruct-MusicGen, we synthesize an instructional dataset using the Slakh2100 dataset~\cite{slakh}, introducing only ~8\% additional parameters compared to the original model, and finetune the model for only 5K steps, which is less than 1\% of training a music editing model from scratch.

We evaluate Instruct-MusicGen on two datasets: the Slakh test set and the out-of-domain MoisesDB dataset~\cite{moisesdb}. Our model outperforms existing baselines and achieves performance comparable to models specifically trained for individual tasks. This demonstrates the effectiveness of our approach in leveraging pretrained models for text-to-music editing while maintaining high-quality results. 


   



\section{Related work}\label{sec:related_work}


Text-based music editing provides a flexible approach for editing music using textual queries. This method is similar to those used in other modalities that require editing, such as image~\cite{instructpix2pix, visualinstruct} and video~\cite{stablevideo, videoedit} editing. In text-to-music editing, text is used to specify precise alterations to existing music compositions. Previous research such as AUDIT~\cite{AUDIT} and InstructME~\cite{instructME} developed a diffusion model trained with paired music editing data. Additionally, models like M$^2$UGen~\cite{M2UGen}, Loop Copilot~\cite{loopcopilot}, MusicAgent~\cite{musicagent}, ComposerX~\cite{composerx} and WavCraft~\cite{wavcraft} use large language models (LLMs) for reasoning and regenerate music with external music generation models. Furthermore, GMSDI~\cite{GMSDI} attempts to model a joint multi-stem distribution of music for text-based generation and separation. Certain models focus exclusively on specific tasks within music editing, such as conditioned generation~\cite{cocomulla, airgen, musiccontrolnet} and separation~\cite{audiosep}, along with intra-stem editing tasks such as text-based timbre transfer and style transfer~\cite{musicmagus, zeroshotddpm, timevaryinversion, audiopromptadapter}.

The task of inter-stem music editing is closely related to stem-wise music generation. Although not directly tied to text-based controls, some research focuses on modeling stem-wise representations to enable simultaneous stem generation and separation. For instance, Jen-1 Composer~\cite{jen1composer} and MSDM~\cite{msdm} jointly model the distribution of music with four stems using a diffusion model. The abilities of most existing stem-wise music models are restricted to a fixed set of 4 stems, which limits flexibility but enhances controllability. Besides, StemGen~\cite{stemgen} trains a LLaMA-based auto-regressive model for flexible stem-wise audio generation. 

 Our work distinguishes itself from these existing efforts in several key ways. First, rather than developing a new model from scratch or strictly adhering to a fixed set of stems, we leverage the power of a pretrained music language model, MusicGen, and enhance it with instruction tuning. This approach not only reduces the computational cost but also retains the high audio quality of the original MusicGen model. Furthermore, our method introduces minimal additional parameters and requires significantly less training, demonstrating a more efficient and scalable solution for text-based music editing. 


\section{Method}\label{sec:method}

\subsection{MusicGen}

The original MusicGen consists of three components: (1) the EnCodec~\cite{encodec} audio encoder and decoder, which compress music audio waveforms into latent codes and reconstruct them back into waveforms; (2) a multi-layer transformer architecture that models sequences of latent codes, capturing higher-level music representations and efficiently modeling internal relationships within music audio; and (3) the T5~\cite{t5} text encoder, which converts text descriptions into embeddings for text-conditioned generation.

EnCodec employs Residual Vector Quantization (RVQ)~\cite{rvq} to compress audio into tokens using multiple codebooks, where each quantizer encodes the quantization error from the previous one. For a reference audio $X \in \mathbb{R}^{d \cdot f_n}$, where $d$ is the duration and $f_n$ is the sample rate, EnCodec compresses $X$ into $Q \in \{1, ..., L\}^{N \times d \cdot f_s}$, where $L$ is the RVQ codebook size, $N$ is the number of codebooks, and $f_s$ is the latent code sample rate ($f_s << f_n$). In MusicGen, $N=4$, $f_n = 50$, $f_s = 32000$, and $L = 2048$. Finally, the transformer models the sequence relationships over latent codes\footnote{MusicGen's encodec uses a 50Hz sample rate, which is different from the original 75Hz EnCodec model.}. 




\subsection{Instruct-MusicGen}

MusicGen is a text-to-music generation model, capable of generating music audio from a given text prompt. However, MusicGen cannot edit existing music audio. To address this limitation, we introduce Instruct-MusicGen, which transforms MusicGen into a model that can follow editing instructions to modify existing music audio.

Instruct-MusicGen takes a music audio input $X^{\text{cond}}$ and a text instruction $X^{\text{instruct}}$ (e.g., "Add guitar") as inputs. The model then edits the music audio $X^{\text{cond}}$ according to the instruction $X^{\text{instruct}}$ and generates the desired edited music $X^{\text{music}}$. As illustrated in Figure~\ref{fig:2}, Instruct-MusicGen incorporates two additional modules into the vanilla MusicGen: an audio fusion and a text fusion module. 

\begin{figure}[t]
    \centering
\includegraphics[width=\linewidth]{figs/diagram_v6.pdf}
    \caption{Illustration of the fusion mechanism inside the Transformer module of instruct-MusicGen. The audio fusion module transforms the conditional music audio into embeddings using a duplicated encoder and integrates these embeddings into the MusicGen decoder. The text fusion module modifies the cross-attention mechanism to handle text instructions by finetuning specific layers while keeping the text encoder parameters frozen. }
    \label{fig:2}
\end{figure}

\subsubsection{Audio Fusion Module}

The audio fusion module enables Instruct-MusicGen to accept external audio inputs, which is inspired by LLaMA-Adapter~\cite{llamaadapter} and Coco-mulla~\cite{cocomulla}. The lower part of Figure~\ref{fig:2} illustrates the audio fusion module. Initially, we convert $X^{\text{cond}}$ into EnCodec tokens, followed by re-encoding these tokens into the embedding $z^{\text{cond}}$ through the pre-trained embedding layers of MusicGen. Similarly, we transform $X^{\text{music}}$ into the pretrained embedding $z^{\text{music}}$.

The module begins with duplicating self-attention modules of the pretrained MusicGen model to extract latent representations of $z^{\text{cond}}$. Given that MusicGen consists of $M$ layers, we denote 

\begin{equation}
    Z^{\text{cond}} = \{z^{\text{cond}}_0, z^{\text{cond}}_1, \ldots, z^{\text{cond}}_M\},
\end{equation}

\begin{equation}
    Z^{\text{music}} = \{z^{\text{music}}_0, z^{\text{music}}_1, \ldots, z^{\text{music}}_M\},
\end{equation}

\noindent which represent the hidden states of $X^{\text{cond}}$ and $X^{\text{music}}$ respectively. Note that we use a learnable input embedding as $z^{\text{cond}}_0$ and initialize $z^{\text{music}}_0$ with $z^{\text{music}}$. 

We compute the vanilla self attention for $X^{\text{music}}$ as follows:
\begin{equation}
    Q^{\text{music}}_l, K^{\text{music}}_l, V^{\text{music}}_l = \text{QKV-projector}(z^{\text{music}}_l),
\end{equation}
\begin{equation}
    o^{\text{music}}_l = \text{SelfAttn}(Q^{\text{music}}_l, K^{\text{music}}_l, V^{\text{music}}_l).
\end{equation}

We project $z^{\text{cond}}$ to a high-dimension representation $h$ via a linear layer $f_l$ and learnable positional encoding $e_l$,
\begin{equation}
h=f_l(z^{\text{cond}}) + e_l.
\end{equation}
Then, we compute the $(l+1)$-th layer hidden states of $X^{\text{cond}}$ as follows:
\begin{equation}
    Q^{\text{cond}}_l, K^{\text{cond}}_l, V^{\text{cond}}_l = \text{QKV-projector}(z^{\text{cond}}_l + h),
\end{equation}
\begin{equation}
   z^{\text{cond}}_{l+1} = {\rm SelfAttn}(Q^{\text{cond}}_l, K^{\text{cond}}_l, V^{\text{cond}}_l).
\end{equation}


To fuse information of $X^{\text{cond}}$ into $X^{\text{music}}$, we compute the cross attention between them,
\begin{equation}
    s^{\text{music}}_l = \text{CrossAttn}(Q^{\text{music}}_l+Q^{\text{cond}}_l, K^{\text{cond}}_l, V^{\text{cond}}_l).
\end{equation}

Finally, the attention output of $X^{\text{music}}$ is updated as follows,
\begin{equation}
s'_{l} = o^{\text{music}}_l + g_l \cdot s^{\text{music}}_l,
\end{equation}
\begin{equation}
z^{\text{music}}_{l+1} = \text{TextFusion}(s'_{l}, X^{\text{instruct}}),
\label{eq:textfusion}
\end{equation}
where $g$ is a zero-initialized learnable gating factor.

Thus, the total trainable parameters in Instruct-MusicGen include the input embedding $z^{\rm cond}_0$, linear layers $f_l$, learnable position embeddings $e_l$, learnable gating factors $g$, and learnable parameters in the text fusion module.

\subsubsection{Text Fusion Module}

To replace the text description input with instruction input, we modify the behavior of the current text encoder.
We achieve this by finetuning only the cross-attention module between the text embedding and the music representations while keeping the text encoder's parameters frozen.

The instruction is embedded and encoded by the T5 text encoder as $z^{\text{instruct}} = \text{T5}(X^{\text{instruct}})$. 
For efficient finetuning of the cross-attention module, we apply LoRA to the query and value projection layers. Thus, we expand Equation~\ref{eq:textfusion} as follows,
\begin{equation}
    Q_l, K^{\text{instruct}}_l, V^{\text{instruct}}_l = \text{QKV-Lora}(s'_l, z^{\text{instruct}}),
\end{equation}
\begin{equation}
    z^{\text{music}}_{l+1} = \text{CrossAttn}(Q_l, K^{\text{instruct}}_l, V^{\text{instruct}}_l).
\end{equation}
During fine-tuning, only query and value projection layers are trainable in the text fusion module.

\section{Experiments}\label{sec:experiment}

\begin{figure}[tbp]
    \centering
    \subfigure[Input music.]{
        \includegraphics[width=0.4\textwidth]{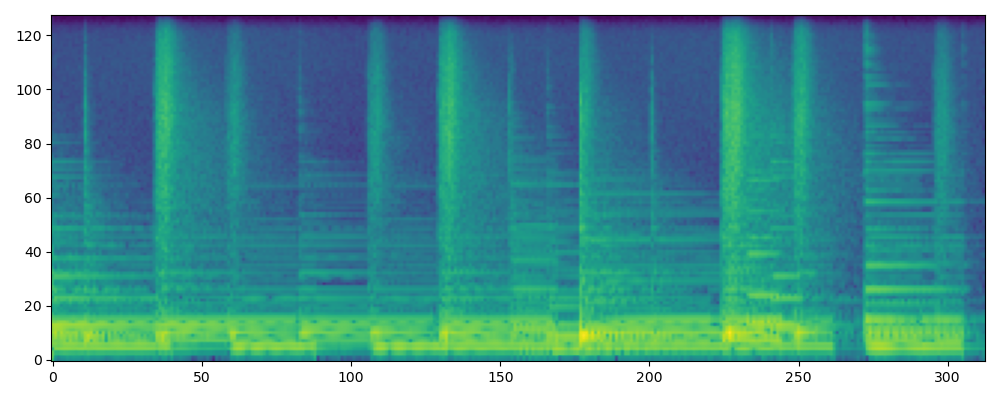}
    }
    \subfigure[Edited music output.]{
        \includegraphics[width=0.4\textwidth]{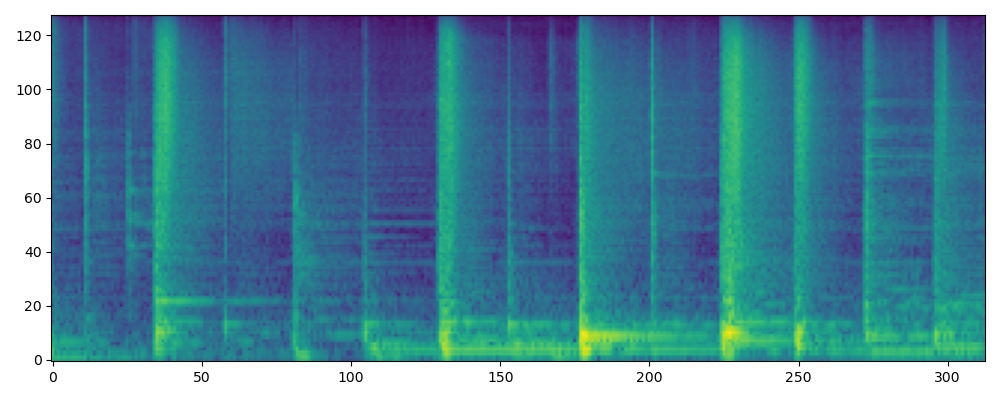}
    }
    \subfigure[Ground truth.]{
        \includegraphics[width=0.4\textwidth]{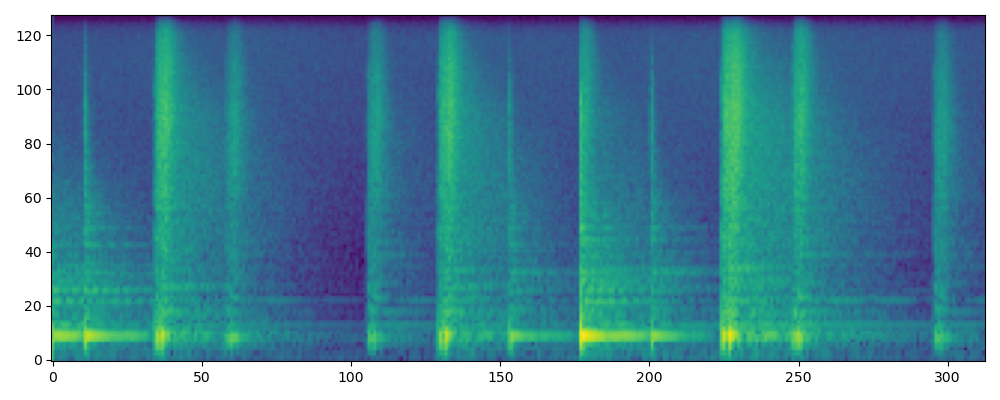}
    }
    \caption{Spectrograms when Instruct-MusicGen removes the drum stem. }
    \label{fig:subfigures}
\end{figure}

We conduct both subjective experiments and objective experiments for evaluation, and also provide example spectrograms in Figure~\ref{fig:subfigures}.

\subsection{Objective Experiments}

\subsubsection{Dataset}

For our objective evaluations, we utilise two distinct datasets, each serving a specific purpose in assessing both in-domain and out-of-domain performance capabilities of various models.

\begin{enumerate}[itemsep=0pt, parsep=0.6pt]
    \item \textbf{Slakh2100 dataset}~\cite{slakh}. The Synthesized Lakh (Slakh) Dataset, originally derived from the Lakh MIDI Dataset v0.1, comprises audio tracks synthesised using high-quality sample-based virtual instruments. This dataset features 2100 tracks complete with corresponding MIDI files. 
    \item \textbf{MoisesDB dataset}~\cite{moisesdb}. The MoisesDB dataset includes 240 real audio tracks sourced from 45 diverse artists spanning twelve musical genres. Uniquely, MoisesDB organises its tracks into a detailed two-level hierarchical taxonomy of stems, offering a varied number of stems per track, each annotated with textual descriptions. 
\end{enumerate}

The rationale for selecting two datasets lies in their diverse configurations and common applications. While the Slakh dataset is traditionally utilised for training models tailored to a four-stem arrangement, our model, Instruct-MusicGen, although initially trained on this dataset, is designed to generalise to various stem configurations. Conversely, models such as InstructME and AUDIT are trained on private or larger, more diverse datasets. By employing both Slakh2100 and MoisesDB, we ensure a comprehensive evaluation, allowing us to fairly compare the adaptability and performance of different models under varying conditions of data familiarity and complexity.

\subsubsection{Data Preprocessing}

We utilised the Slakh2100 dataset to construct an instruction-based dataset for our experiments, employing the following pipeline:

\begin{itemize}[itemsep=0pt, parsep=0.6pt]
    
\item A data point was randomly selected from the Slakh training dataset.
\item An instruction was chosen from a predefined set \textit{\{add, remove, extract\}} along with a target stem. Subsequently, $n$ other stems were selected from the remaining stems.
\item An offset was randomly determined to cut a 5-second audio clip. If the target stem contained more than 50\% silence, a different offset was selected.
\item The stems were mixed according to the specified instructions to create a triplet consisting of \textit{\{instruction text, condition audio input, audio ground truth\}}.
\end{itemize}

\subsubsection{Experimental Setup}

For the finetuning of MusicGen, we jointly trained the audio fusion module and the text fusion module. The optimisation process utilised the AdamW optimiser, with a learning rate set at $5\times10^{-3}$. We use L2 loss over latent tokens as the training objective. Training incorporated a Cosine Annealing scheduler with an initial warmup of 100 steps. The training regimen extended over 5,000 steps with an accumulated batch size of 32, achieved through setting the batch size to 8 and using gradient accumulation over 4 iterations. The finetuning process was executed on a single NVIDIA A100 GPU and was completed within two days.

\subsubsection{Baselines}
In this section, we explore two baseline models, each distinguished by their unique methodologies for handling audio data. 

\begin{enumerate}[itemsep=0pt, parsep=0.6pt]
    \item \textbf{AUDIT}~\cite{AUDIT}: AUDIT is an instruction-guided audio editing model, consisting of a variational autoencoder (VAE) for converting input audio into a latent space representation, a T5 text encoder for processing edit instructions, and a diffusion network that performs the actual audio editing in the latent space. The system accepts mel-spectrograms of input audio and edit instructions, and generates the edited audio as output. 
    

    \item \textbf{M$^2$UGen}~\cite{M2UGen}: The M$^2$UGen framework leverages large language models to comprehend and generate music across various modalities, integrating abilities from external models such as MusicGen~\cite{musicgen} and AudioLDM 2~\cite{audioldm2}. It is designed to stimulate creative outputs from diverse sources, showcasing robust performance in multi-modal music generation. 

\end{enumerate}

Besides, InstructME can also perform instruction-guided music editing and remixing with latent diffusion models. We exclude it from comparison because InstructME's model weights and evaluation protocol are not publicly released.



\begin{table}[htbp]
\small
    \centering
    \begin{tabular}{ccccc}
    \toprule
        \textbf{Model}  & \textbf{Param size} & \textbf{Dataset} & \textbf{Hours (h)} & \textbf{Steps}\\
    \midrule
        AUDIT  & 942M (1.5B) & Multiple & $\sim$6500 & 0.5M\\
        InstructME & 967M (1.7B) & Multiple & 417 & 2M \\
        M$^2$UGen  & 637M ($\sim$9B) & MUEdit & 60.22 & - \\
    \midrule
        \textbf{Ours} & \textbf{264M} (3.5B) & Slakh & 145 & \textbf{5K} \\
     \bottomrule
    \end{tabular}
    \caption{Comparison of different models, where the param size numbers are trainable parameters and total parameters respectively. Our model has the lowest parameter size, and only requires 5K training steps.}
    \label{tab:baselines}
\end{table}

\subsubsection{Metrics}

The metrics to evaluate model performance are listed below.

\begin{enumerate}[itemsep=0pt, parsep=0.6pt]

\item \textbf{Fr\'{e}chet Audio Distance (FAD)}~\cite{fad}\footnote{\url{https://github.com/gudgud96/frechet-audio-distance}.} measures the similarity between two sets of audio files by comparing multivariate Gaussian distributions fitted to feature embeddings from the audio data. We use the FAD score to evaluate the overall audio quality of the predicted music.

\item \textbf{CLAP Score (CLAP)}~\cite{clap}\footnote{\url{https://github.com/LAION-AI/CLAP}.} is used in our experiments to measure the correspondence between the edited music and a target text. For the removal task, the target text is generated by deleting the name of the removed instrument from the original text.

\item \textbf{Kullback-Leibler Divergence (KL)}\footnote{\url{https://github.com/haoheliu/audioldm_eval}.} assesses the difference between the probability distributions of audio features from two sources, indicating information loss when approximating one distribution with another. A low KL score indicates the predicted music shares similar features with the ground truth.

\item \textbf{Structural Similarity (SSIM)}~\cite{ssim} is an image quality metric that we adapt to evaluate structural similarity between predicted music and ground truth.

\item \textbf{Scale-Invariant Signal-to-Distortion Ratio (SI-SDR)}~\cite{si-sdr} quantifies audio quality, especially in source separation tasks. It is scale-invariant, useful for varying audio volumes, and measures distortion relative to a reference signal. We use SI-SDR to evaluate the signal loss of the predicted audio.

\item \textbf{Scale-Invariant Signal-to-Distortion Ratio improvement (SI-SDRi)}~\cite{si-sdri} extends SI-SDR, measuring the improvement in signal-to-distortion ratio before and after processing. It is commonly used in audio enhancement and separation contexts.

\end{enumerate}

To further investigate whether the model successfully adds, removes or extracts the instrument, we propose the \textbf{P-Demucs score} to evaluate the model performance. This metric specifically focuses on detecting the presence of a newly added instrument in the generated audio. It leverages the Demucs model, a source separation model, to isolate the target instrument from the audio. After separation, the root-mean-square energy (RMSE) of the isolated track is analyzed. For example, if the instruction is to "add guitar," a non-silent guitar track is regarded as a successful edit.

\subsubsection{Objective Experiment Results}

Our evaluation of Instruct-MusicGen demonstrates its superior performance across various tasks compared to existing text-to-music editing baselines (AUDIT, InstructME, M$^2$UGen). On the Slakh dataset (Table~\ref{tab:editing_4_stems}), Instruct-MusicGen excelled in adding, removing, and extracting stems, achieving the lowest Fréchet Audio Distance (FAD) and highest CLAP and SSIM scores. It also significantly improved the signal-to-noise ratio (SI-SDR) in the removal task, showing balanced performance across all metrics and proving its robustness in various editing scenarios. 
Similarly, in the MoisesDB dataset evaluations (Table~\ref{tab:moisesDB}), Instruct-MusicGen demonstrated strong performance, with the best performance on most metrics over the three tasks.

\begin{table*}[t]
\small
    \centering
    \begin{tabular}{c|l|ccccccc}
    \toprule
    \toprule
    \textbf{Task} & \textbf{Models}  & \textbf{FAD}$\downarrow$ & \textbf{CLAP}$\uparrow$ & \textbf{KL}$\downarrow$ & \textbf{SSIM}$\uparrow$ &\textbf{P-Demucs}$\uparrow$& \textbf{SI-SDR}$\uparrow$ & \textbf{SI-SDRi}$\uparrow$ \\
    \midrule
    \midrule
    \multirow{5}{*}{\textbf{Add}} & AUDIT & 6.88 & 0.12  & 1.02   & 0.21 &0.53& - & - \\
    & M$^2$UGen  & 7.24 & 0.22 & 0.99  & 0.20 &0.43 & - & - \\
    \cmidrule{2-9}
    \cmidrule{2-9}
    & \textbf{Ours}  & \textbf{3.75} & \textbf{0.23}  & \textbf{0.67}   & \textbf{0.26}  & \textbf{0.80} & - & - \\
    \midrule
    \midrule
    \multirow{3}{*}{\textbf{Remove}} & AUDIT  & 15.48 & 0.07  & 2.75  & 0.35 & 0.33  & -45.60  & -47.28  \\
    & M$^2$UGen  & 8.26 & 0.09 & 1.59 & 0.23 & 0.70  & -44.20 &  -46.13 \\
    \cmidrule{2-9}
    \cmidrule{2-9}
    & \textbf{Ours}  & \textbf{3.35} & \textbf{0.12}  & \textbf{0.66}  & \textbf{0.45} & \textbf{0.76}  & \textbf{-2.09}  & \textbf{-3.77}   \\
    \midrule
    \midrule
    \multirow{3}{*}{\textbf{Extract}} & AUDIT  & 15.08 & 0.06  & 2.38   & 0.42 & 0.61 & -52.90  & -50.16  \\
    & M$^2$UGen  & 8.14 & 0.11  & 2.15  & 0.31 & 0.60 & -46.38 & -43.53 \\
    \cmidrule{2-9}
    \cmidrule{2-9}
    & \textbf{Ours} & \textbf{3.24} & \textbf{0.12}  & \textbf{0.54}  & \textbf{0.52} & \textbf{0.75}  & \textbf{-9.00}  & \textbf{-6.15}  \\    
    \bottomrule
    \bottomrule
    \end{tabular}
    \caption{Comparison of text-based music editing models on the Slakh dataset (4 stems). }
    \label{tab:editing_4_stems}
\end{table*}

\begin{table*}[ht]
\small
    \centering
    \begin{tabular}{c|l|cccccccc}
    \toprule
    \toprule
    \textbf{Task} & \textbf{Models} & \textbf{FAD}$\downarrow$ & \textbf{CLAP}$\uparrow$ & \textbf{KL}$\downarrow$ & \textbf{SSIM}$\uparrow$ &\textbf{P-Demucs}$\uparrow$& \textbf{SI-SDR}$\uparrow$ & \textbf{SI-SDRi}$\uparrow$ \\
    \midrule
    \midrule
    \multirow{3}{*}{\textbf{Add}} & AUDIT  & 4.06 & 0.12  & 0.84   & 0.21 &0.50 & - & - \\
    & M$^2$UGen  & 5.00 & \textbf{0.18} & 0.83  & 0.20 &0.45& - & - \\
    \cmidrule{2-9}
    & \textbf{Ours}  & \textbf{3.79} & \textbf{0.18} & \textbf{0.35}  & \textbf{0.35} &\textbf{0.77}& - & - \\
    \midrule
    \midrule
    \multirow{3}{*}{\textbf{Remove}} & AUDIT  & 10.72 & 0.10 & 2.46 & \textbf{0.34} &0.41 & -44.32 & -57.10 \\
    & M$^2$UGen  & \textbf{3.75} & \textbf{0.13} & 1.27 & 0.19 &0.72& -43.94 & -56.73 \\
    \cmidrule{2-9}
    & \textbf{Ours}  & 5.05 & 0.10 & \textbf{0.84} & \textbf{0.34}& \textbf{0.78} & \textbf{-13.70} & \textbf{-26.48}   \\
    \midrule
    \midrule
    \multirow{3}{*}{\textbf{Extract}} & AUDIT  & 6.67 & 0.07 & 1.97 & \textbf{0.45} &0.60 & -54.53 & -56.17 \\
    & M$^2$UGen  & 5.74 & 0.08 & 1.91 & 0.25 & 0.52 & -42.84 & -44.49 \\
    \cmidrule{2-9}
    & \textbf{Ours}  & \textbf{4.96} & \textbf{0.11} & \textbf{1.36} & 0.40 & \textbf{0.78} & \textbf{-21.39} & \textbf{-23.03}  \\    
    \bottomrule
    \bottomrule
    \end{tabular}
    \caption{Comparison of text-based music editing models on the MoisesDB dataset.}
    \label{tab:moisesDB}
\end{table*}

We find that all models exhibit negative SI-SDR and SI-SDRi scores, which is a common occurrence when evaluating generative models on a signal level. These metrics are typically designed for source separation tasks and are not entirely fair to generative models, as they penalise even minor discrepancies between the generated and original signals. Generative models like Instruct-MusicGen often focus on producing perceptually plausible audio rather than perfectly matching the original signal at a technical level.


\subsection{Subjective Experiments}

\subsubsection{Experimental Setup}

We conducted a subjective listening test to evaluate the model's performance.\footnote{This subjective test was approved by the ethics committee of Sony.} This test involved disseminating an online survey within the Music Information Retrieval (MIR) community and our broader research network, which resulted in the collection of 30 complete responses. The gender distribution of the participants was 23 males (76.7\%) and 7 females (23.3\%). Regarding professional musical education experience, 4 participants (13.3\%) had less than 1 year of experience, 13 (43.3\%) had between 1 and 5 years, and 13 participants (43.3\%) had more than 5 years of experience.  For the data preparation, we randomly selected a subset of data points from the objective test dataset. Specifically, 6 audio samples were chosen, comprising 2 audio samples for each subtask (add, remove, extract). Each data point included results from the baseline models, our models, and the ground truth from the dataset. 

\subsubsection{Metrics}

\begin{enumerate}[itemsep=0pt, parsep=0.6pt]
    \item \textbf{Instruction Adherence (IA)} assesses how accurately the generated music follows the given editing instruction. In this experiment, participants rate the generated music on a scale from 1 to 5, where 1 indicates that the instruction was not followed at all, and 5 indicates that the instruction was followed perfectly. For example, if the instruction is "Remove Drums," a rating of 1 would mean that the drums were not removed at all, while a rating of 5 would mean that the drums were completely removed.

\item \textbf{Audio Quality (AQ)} evaluates the overall audio quality of the generated music in comparison to the original music. Participants rate the audio quality on a scale from 1 to 5, where 1 represents very poor quality with significant degradation compared to the original music, and 5 represents excellent quality, as good as or better than the original music. This metric helps in understanding how the editing process affects the overall sound quality of the music.
\end{enumerate}

\subsubsection{Subjective Experiment Results}

\begin{table}[ht]
\footnotesize
    \centering
    \begin{tabular}{l|cc}
    \toprule
        \textbf{Model} & \textbf{Instruction Adherence}$\uparrow$ & \textbf{Audio Quality}$\uparrow$ \\
        \midrule
        AUDIT & 1.54 &	2.56 \\
        M$^2$UGen & 1.70 &	1.92 \\
        \midrule
        \textbf{Ours} & \textbf{3.85} & \textbf{3.55}\\
        \midrule
        Ground truth & 4.36 & 4.21\\
        \bottomrule
    \end{tabular}
    \caption{The subjective experiment results. }
    \label{tab:subjective}
\end{table}

The results of our subjective experiments are summarised in Table \ref{tab:subjective}. We conducted two paired t-tests with Bonferroni correction, setting the significance level at $\alpha=0.05$. The results shows that our model demonstrates a significant improvement in both Instruction Adherence (IA) and Audio Quality (AQ) compared to the baseline models, AUDIT and M$^2$UGen. \footnote{More audio samples can be found at \url{https://bit.ly/instruct-musicgen}.}

\section{Conclusion}\label{sec:conclusion}


In this paper, we introduced Instruct-MusicGen, a novel approach to text-to-music editing that fosters joint musical and textual controls. By finetuning the existing MusicGen model with instruction tuning, Instruct-MusicGen demonstrated its capability of editing music in various ways, including adding, separating and extracting a stem from music audio using textual queries, without the need for training specialised models from scratch. Also, it outperforms various baseline models that are dedicated to specific music editing tasks. Furthermore, our method uses significantly fewer resources than previous models, with a requirement of tuning only 8\% of the parameters of the original MusicGen. 

\section{Acknowledgements}

This work was done during Yixiao Zhang's internship at Sony AI. Yixiao Zhang was a research student at the UKRI Centre for Doctoral Training in Artificial Intelligence and Music, supported jointly by the China Scholarship Council, Queen Mary University of London and Apple Inc.

\bibliography{ISMIRtemplate}

\begin{thebibliography}{10}
\providecommand{\url}[1]{#1}
\csname url@samestyle\endcsname
\providecommand{\newblock}{\relax}
\providecommand{\bibinfo}[2]{#2}
\providecommand{\BIBentrySTDinterwordspacing}{\spaceskip=0pt\relax}
\providecommand{\BIBentryALTinterwordstretchfactor}{4}
\providecommand{\BIBentryALTinterwordspacing}{\spaceskip=\fontdimen2\font plus
\BIBentryALTinterwordstretchfactor\fontdimen3\font minus \fontdimen4\font\relax}
\providecommand{\BIBforeignlanguage}[2]{{%
\expandafter\ifx\csname l@#1\endcsname\relax
\typeout{** WARNING: IEEEtran.bst: No hyphenation pattern has been}%
\typeout{** loaded for the language `#1'. Using the pattern for}%
\typeout{** the default language instead.}%
\else
\language=\csname l@#1\endcsname
\fi
#2}}
\providecommand{\BIBdecl}{\relax}
\BIBdecl

\bibitem{musiclm}
\BIBentryALTinterwordspacing
A.~Agostinelli, T.~I. Denk, Z.~Borsos, J.~H. Engel, M.~Verzetti, A.~Caillon, Q.~Huang, A.~Jansen, A.~Roberts, M.~Tagliasacchi, M.~Sharifi, N.~Zeghidour, and C.~H. Frank, ``Music{LM}: Generating music from text,'' \emph{CoRR}, vol. abs/2301.11325, 2023. [Online]. Available: \url{https://doi.org/10.48550/arxiv.2301.11325}
\BIBentrySTDinterwordspacing

\bibitem{musicgen}
\BIBentryALTinterwordspacing
J.~Copet, F.~Kreuk, I.~Gat, T.~Remez, D.~Kant, G.~Synnaeve, Y.~Adi, and A.~D{\'{e}}fossez, ``Simple and controllable music generation,'' in \emph{Advances in Neural Information Processing Systems 36: Annual Conference on Neural Information Processing Systems 2023, NeurIPS 2023, New Orleans, LA, USA, December 10 - 16, 2023}, A.~Oh, T.~Naumann, A.~Globerson, K.~Saenko, M.~Hardt, and S.~Levine, Eds., 2023. [Online]. Available: \url{http://papers.nips.cc/paper\_files/paper/2023/hash/94b472a1842cd7c56dcb125fb2765fbd-Abstract-Conference.html}
\BIBentrySTDinterwordspacing

\bibitem{jen1}
\BIBentryALTinterwordspacing
P.~Li, B.~Chen, Y.~Yao, Y.~Wang, A.~Wang, and A.~Wang, ``{JEN-1}: Text-guided universal music generation with omnidirectional diffusion models,'' \emph{CoRR}, vol. abs/2308.04729, 2023. [Online]. Available: \url{https://doi.org/10.48550/arxiv.2308.04729}
\BIBentrySTDinterwordspacing

\bibitem{audioldm2}
\BIBentryALTinterwordspacing
H.~Liu, Q.~Tian, Y.~Yuan, X.~Liu, X.~Mei, Q.~Kong, Y.~Wang, W.~Wang, Y.~Wang, and M.~D. Plumbley, ``Audio{LDM} 2: Learning holistic audio generation with self-supervised pretraining,'' \emph{CoRR}, vol. abs/2308.05734, 2023. [Online]. Available: \url{https://doi.org/10.48550/arxiv.2308.05734}
\BIBentrySTDinterwordspacing

\bibitem{musicldm}
\BIBentryALTinterwordspacing
K.~Chen, Y.~Wu, H.~Liu, M.~Nezhurina, T.~Berg{-}Kirkpatrick, and S.~Dubnov, ``Music{LDM}: Enhancing novelty in text-to-music generation using beat-synchronous mixup strategies,'' \emph{CoRR}, vol. abs/2308.01546, 2023. [Online]. Available: \url{https://doi.org/10.48550/arxiv.2308.01546}
\BIBentrySTDinterwordspacing

\bibitem{cocomulla}
\BIBentryALTinterwordspacing
L.~Lin, G.~Xia, J.~Jiang, and Y.~Zhang, ``Content-based controls for music large language modeling,'' \emph{CoRR}, vol. abs/2310.17162, 2023. [Online]. Available: \url{https://doi.org/10.48550/arxiv.2310.17162}
\BIBentrySTDinterwordspacing

\bibitem{musiccontrolnet}
S.-L. Wu, C.~Donahue, S.~Watanabe, and N.~J. Bryan, ``Music controlnet: Multiple time-varying controls for music generation,'' \emph{IEEE/ACM Transactions on Audio, Speech, and Language Processing}, vol.~32, pp. 2692--2703, 2024.

\bibitem{mustango}
J.~Melechovsky, Z.~Guo, D.~Ghosal, N.~Majumder, D.~Herremans, and S.~Poria, ``Mustango: Toward controllable text-to-music generation,'' \emph{arxiv preprint arxiv:2311.08355}, 2023.

\bibitem{airgen}
\BIBentryALTinterwordspacing
L.~Lin, G.~Xia, Y.~Zhang, and J.~Jiang, ``Arrange, inpaint, and refine: Steerable long-term music audio generation and editing via content-based controls,'' \emph{CoRR}, vol. abs/2402.09508, 2024. [Online]. Available: \url{https://doi.org/10.48550/arxiv.2402.09508}
\BIBentrySTDinterwordspacing

\bibitem{instructME}
\BIBentryALTinterwordspacing
B.~Han, J.~Dai, X.~Song, W.~Hao, X.~He, D.~Guo, J.~Chen, Y.~Wang, and Y.~Qian, ``Instruct{ME}: An instruction guided music edit and remix framework with latent diffusion models,'' \emph{CoRR}, vol. abs/2308.14360, 2023. [Online]. Available: \url{https://doi.org/10.48550/arxiv.2308.14360}
\BIBentrySTDinterwordspacing

\bibitem{musicmagus}
\BIBentryALTinterwordspacing
Y.~Zhang, Y.~Ikemiya, G.~Xia, N.~Murata, M.~A.~M. Ram{\'{\i}}rez, W.~Liao, Y.~Mitsufuji, and S.~Dixon, ``Music{M}agus: Zero-shot text-to-music editing via diffusion models,'' \emph{CoRR}, vol. abs/2402.06178, 2024. [Online]. Available: \url{https://doi.org/10.48550/arxiv.2402.06178}
\BIBentrySTDinterwordspacing

\bibitem{M2UGen}
\BIBentryALTinterwordspacing
A.~S. Hussain, S.~Liu, C.~Sun, and Y.~Shan, ``M\({}^{\mbox{2}}\){UG}en: Multi-modal music understanding and generation with the power of large language models,'' \emph{CoRR}, vol. abs/2311.11255, 2023. [Online]. Available: \url{https://doi.org/10.48550/arxiv.2311.11255}
\BIBentrySTDinterwordspacing

\bibitem{AUDIT}
\BIBentryALTinterwordspacing
Y.~Wang, Z.~Ju, X.~Tan, L.~He, Z.~Wu, J.~Bian, and S.~Zhao, ``{AUDIT}: Audio editing by following instructions with latent diffusion models,'' in \emph{Advances in Neural Information Processing Systems 36: Annual Conference on Neural Information Processing Systems 2023, NeurIPS 2023, New Orleans, LA, USA, December 10 - 16, 2023}, A.~Oh, T.~Naumann, A.~Globerson, K.~Saenko, M.~Hardt, and S.~Levine, Eds., 2023. [Online]. Available: \url{http://papers.nips.cc/paper\_files/paper/2023/hash/e1b619a9e241606a23eb21767f16cf81-Abstract-Conference.html}
\BIBentrySTDinterwordspacing

\bibitem{loopcopilot}
\BIBentryALTinterwordspacing
Y.~Zhang, A.~Maezawa, G.~Xia, K.~Yamamoto, and S.~Dixon, ``Loop {C}opilot: Conducting {AI} ensembles for music generation and iterative editing,'' \emph{CoRR}, vol. abs/2310.12404, 2023. [Online]. Available: \url{https://doi.org/10.48550/arxiv.2310.12404}
\BIBentrySTDinterwordspacing

\bibitem{uniaudio}
\BIBentryALTinterwordspacing
D.~Yang, J.~Tian, X.~Tan, R.~Huang, S.~Liu, X.~Chang, J.~Shi, S.~Zhao, J.~Bian, X.~Wu, Z.~Zhao, S.~Watanabe, and H.~Meng, ``Uni{A}udio: An audio foundation model toward universal audio generation,'' \emph{CoRR}, vol. abs/2310.00704, 2023. [Online]. Available: \url{https://doi.org/10.48550/arxiv.2310.00704}
\BIBentrySTDinterwordspacing

\bibitem{llamaadapter}
\BIBentryALTinterwordspacing
R.~Zhang, J.~Han, A.~Zhou, X.~Hu, S.~Yan, P.~Lu, H.~Li, P.~Gao, and Y.~Qiao, ``{LLaMA-Adapter}: Efficient fine-tuning of language models with zero-init attention,'' \emph{CoRR}, vol. abs/2303.16199, 2023. [Online]. Available: \url{https://doi.org/10.48550/arxiv.2303.16199}
\BIBentrySTDinterwordspacing

\bibitem{lora}
\BIBentryALTinterwordspacing
E.~J. Hu, Y.~Shen, P.~Wallis, Z.~Allen{-}Zhu, Y.~Li, S.~Wang, and W.~Chen, ``{LoRA}: Low-rank adaptation of large language models,'' \emph{CoRR}, vol. abs/2106.09685, 2021. [Online]. Available: \url{https://arxiv.org/abs/2106.09685}
\BIBentrySTDinterwordspacing

\bibitem{slakh}
\BIBentryALTinterwordspacing
E.~Manilow, G.~Wichern, P.~Seetharaman, and J.~L. Roux, ``Cutting music source separation some {S}lakh: {A} dataset to study the impact of training data quality and quantity,'' in \emph{2019 {IEEE} Workshop on Applications of Signal Processing to Audio and Acoustics, {WASPAA} 2019, New Paltz, NY, USA, October 20-23, 2019}.\hskip 1em plus 0.5em minus 0.4em\relax {IEEE}, 2019, pp. 45--49. [Online]. Available: \url{https://doi.org/10.1109/WASPAA.2019.8937170}
\BIBentrySTDinterwordspacing

\bibitem{moisesdb}
\BIBentryALTinterwordspacing
I.~Pereira, F.~Ara{\'{u}}jo, F.~Korzeniowski, and R.~Vogl, ``Moises{DB}: {A} dataset for source separation beyond 4-stems,'' in \emph{Proceedings of the 24th International Society for Music Information Retrieval Conference, {ISMIR} 2023, Milan, Italy, November 5-9, 2023}, A.~Sarti, F.~Antonacci, M.~Sandler, P.~Bestagini, S.~Dixon, B.~Liang, G.~Richard, and J.~Pauwels, Eds., 2023, pp. 619--626. [Online]. Available: \url{https://doi.org/10.5281/zenodo.10265363}
\BIBentrySTDinterwordspacing

\bibitem{instructpix2pix}
\BIBentryALTinterwordspacing
T.~Brooks, A.~Holynski, and A.~A. Efros, ``Instruct{P}ix2{P}ix: Learning to follow image editing instructions,'' in \emph{{IEEE/CVF} Conference on Computer Vision and Pattern Recognition, {CVPR} 2023, Vancouver, BC, Canada, June 17-24, 2023}.\hskip 1em plus 0.5em minus 0.4em\relax {IEEE}, 2023, pp. 18\,392--18\,402. [Online]. Available: \url{https://doi.org/10.1109/CVPR52729.2023.01764}
\BIBentrySTDinterwordspacing

\bibitem{visualinstruct}
\BIBentryALTinterwordspacing
H.~Liu, C.~Li, Q.~Wu, and Y.~J. Lee, ``Visual instruction tuning,'' in \emph{Advances in Neural Information Processing Systems 36: Annual Conference on Neural Information Processing Systems 2023, NeurIPS 2023, New Orleans, LA, USA, December 10 - 16, 2023}, A.~Oh, T.~Naumann, A.~Globerson, K.~Saenko, M.~Hardt, and S.~Levine, Eds., 2023. [Online]. Available: \url{http://papers.nips.cc/paper\_files/paper/2023/hash/6dcf277ea32ce3288914faf369fe6de0-Abstract-Conference.html}
\BIBentrySTDinterwordspacing

\bibitem{stablevideo}
\BIBentryALTinterwordspacing
W.~Chai, X.~Guo, G.~Wang, and Y.~Lu, ``Stable{V}ideo: Text-driven consistency-aware diffusion video editing,'' in \emph{{IEEE/CVF} International Conference on Computer Vision, {ICCV} 2023, Paris, France, October 1-6, 2023}.\hskip 1em plus 0.5em minus 0.4em\relax {IEEE}, 2023, pp. 22\,983--22\,993. [Online]. Available: \url{https://doi.org/10.1109/ICCV51070.2023.02106}
\BIBentrySTDinterwordspacing

\bibitem{videoedit}
\BIBentryALTinterwordspacing
D.~Ceylan, C.~P. Huang, and N.~J. Mitra, ``Pix2{V}ideo: Video editing using image diffusion,'' in \emph{{IEEE/CVF} International Conference on Computer Vision, {ICCV} 2023, Paris, France, October 1-6, 2023}.\hskip 1em plus 0.5em minus 0.4em\relax {IEEE}, 2023, pp. 23\,149--23\,160. [Online]. Available: \url{https://doi.org/10.1109/ICCV51070.2023.02121}
\BIBentrySTDinterwordspacing

\bibitem{musicagent}
\BIBentryALTinterwordspacing
D.~Yu, K.~Song, P.~Lu, T.~He, X.~Tan, W.~Ye, S.~Zhang, and J.~Bian, ``Music{A}gent: An {AI} agent for music understanding and generation with large language models,'' in \emph{Proceedings of the 2023 Conference on Empirical Methods in Natural Language Processing, {EMNLP} 2023 - System Demonstrations, Singapore, December 6-10, 2023}, Y.~Feng and E.~Lefever, Eds.\hskip 1em plus 0.5em minus 0.4em\relax Association for Computational Linguistics, 2023, pp. 246--255. [Online]. Available: \url{https://doi.org/10.18653/v1/2023.emnlp-demo.21}
\BIBentrySTDinterwordspacing

\bibitem{composerx}
Q.~Deng, Q.~Yang, R.~Yuan, Y.~Huang, Y.~Wang, X.~Liu, Z.~Tian, J.~Pan, G.~Zhang, H.~Lin \emph{et~al.}, ``Composer{X}: Multi-agent symbolic music composition with {LLM}s,'' \emph{arxiv preprint arxiv:2404.18081}, 2024.

\bibitem{wavcraft}
J.~Liang, H.~Zhang, H.~Liu, Y.~Cao, Q.~Kong, X.~Liu, W.~Wang, M.~D. Plumbley, H.~Phan, and E.~Benetos, ``Wav{C}raft: Audio editing and generation with large language models,'' in \emph{ICLR 2024 Workshop on Large Language Model (LLM) Agents}, 2024.

\bibitem{GMSDI}
\BIBentryALTinterwordspacing
E.~Postolache, G.~Mariani, L.~Cosmo, E.~Benetos, and E.~Rodol{\`{a}}, ``Generalized multi-source inference for text conditioned music diffusion models,'' \emph{CoRR}, vol. abs/2403.11706, 2024. [Online]. Available: \url{https://doi.org/10.48550/arxiv.2403.11706}
\BIBentrySTDinterwordspacing

\bibitem{audiosep}
\BIBentryALTinterwordspacing
X.~Liu, Q.~Kong, Y.~Zhao, H.~Liu, Y.~Yuan, Y.~Liu, R.~Xia, Y.~Wang, M.~D. Plumbley, and W.~Wang, ``Separate anything you describe,'' \emph{CoRR}, vol. abs/2308.05037, 2023. [Online]. Available: \url{https://doi.org/10.48550/arxiv.2308.05037}
\BIBentrySTDinterwordspacing

\bibitem{zeroshotddpm}
\BIBentryALTinterwordspacing
H.~Manor and T.~Michaeli, ``Zero-shot unsupervised and text-based audio editing using {DDPM} inversion,'' \emph{CoRR}, vol. abs/2402.10009, 2024. [Online]. Available: \url{https://doi.org/10.48550/arxiv.2402.10009}
\BIBentrySTDinterwordspacing

\bibitem{timevaryinversion}
\BIBentryALTinterwordspacing
S.~Li, Y.~Zhang, F.~Tang, C.~Ma, W.~Dong, and C.~Xu, ``Music style transfer with time-varying inversion of diffusion models,'' in \emph{Thirty-Eighth {AAAI} Conference on Artificial Intelligence, {AAAI} 2024, Thirty-Sixth Conference on Innovative Applications of Artificial Intelligence, {IAAI} 2024, Fourteenth Symposium on Educational Advances in Artificial Intelligence, {EAAI} 2014, February 20-27, 2024, Vancouver, Canada}, M.~J. Wooldridge, J.~G. Dy, and S.~Natarajan, Eds.\hskip 1em plus 0.5em minus 0.4em\relax {AAAI} Press, 2024, pp. 547--555. [Online]. Available: \url{https://doi.org/10.1609/aaai.v38i1.27810}
\BIBentrySTDinterwordspacing

\bibitem{audiopromptadapter}
F.-D. Tsai, S.-L. Wu, H.~Kim, B.-Y. Chen, H.-C. Cheng, and Y.-H. Yang, ``Audio prompt adapter: Unleashing music editing abilities for text-to-music with lightweight finetuning,'' \emph{arXiv preprint arXiv:2407.16564}, 2024.

\bibitem{jen1composer}
\BIBentryALTinterwordspacing
Y.~Yao, P.~Li, B.~Chen, and A.~Wang, ``{JEN-1} {C}omposer: {A} unified framework for high-fidelity multi-track music generation,'' \emph{CoRR}, vol. abs/2310.19180, 2023. [Online]. Available: \url{https://doi.org/10.48550/arxiv.2310.19180}
\BIBentrySTDinterwordspacing

\bibitem{msdm}
\BIBentryALTinterwordspacing
G.~Mariani, I.~Tallini, E.~Postolache, M.~Mancusi, L.~Cosmo, and E.~Rodol{\`{a}}, ``Multi-source diffusion models for simultaneous music generation and separation,'' \emph{CoRR}, vol. abs/2302.02257, 2023. [Online]. Available: \url{https://doi.org/10.48550/arxiv.2302.02257}
\BIBentrySTDinterwordspacing

\bibitem{stemgen}
\BIBentryALTinterwordspacing
J.~D. Parker, J.~Spijkervet, K.~Kosta, F.~Yesiler, B.~Kuznetsov, J.~Wang, M.~Avent, J.~Chen, and D.~Le, ``Stem{G}en: {A} music generation model that listens,'' \emph{CoRR}, vol. abs/2312.08723, 2023. [Online]. Available: \url{https://doi.org/10.48550/arxiv.2312.08723}
\BIBentrySTDinterwordspacing

\bibitem{encodec}
\BIBentryALTinterwordspacing
A.~D{\'{e}}fossez, J.~Copet, G.~Synnaeve, and Y.~Adi, ``High fidelity neural audio compression,'' \emph{CoRR}, vol. abs/2210.13438, 2022. [Online]. Available: \url{https://doi.org/10.48550/arxiv.2210.13438}
\BIBentrySTDinterwordspacing

\bibitem{t5}
\BIBentryALTinterwordspacing
C.~Raffel, N.~Shazeer, A.~Roberts, K.~Lee, S.~Narang, M.~Matena, Y.~Zhou, W.~Li, and P.~J. Liu, ``Exploring the limits of transfer learning with a unified text-to-text transformer,'' \emph{J. Mach. Learn. Res.}, vol.~21, pp. 140:1--140:67, 2020. [Online]. Available: \url{http://jmlr.org/papers/v21/20-074.html}
\BIBentrySTDinterwordspacing

\bibitem{rvq}
\BIBentryALTinterwordspacing
N.~Zeghidour, A.~Luebs, A.~Omran, J.~Skoglund, and M.~Tagliasacchi, ``Sound{S}tream: An end-to-end neural audio codec,'' \emph{{IEEE} {ACM} Trans. Audio Speech Lang. Process.}, vol.~30, pp. 495--507, 2022. [Online]. Available: \url{https://doi.org/10.1109/TASLP.2021.3129994}
\BIBentrySTDinterwordspacing

\bibitem{fad}
\BIBentryALTinterwordspacing
K.~Kilgour, M.~Zuluaga, D.~Roblek, and M.~Sharifi, ``Fr{\'{e}}chet audio distance: {A} reference-free metric for evaluating music enhancement algorithms,'' in \emph{Interspeech 2019, 20th Annual Conference of the International Speech Communication Association, Graz, Austria, 15-19 September 2019}, G.~Kubin and Z.~Kacic, Eds.\hskip 1em plus 0.5em minus 0.4em\relax {ISCA}, 2019, pp. 2350--2354. [Online]. Available: \url{https://doi.org/10.21437/Interspeech.2019-2219}
\BIBentrySTDinterwordspacing

\bibitem{clap}
\BIBentryALTinterwordspacing
Y.~Wu, K.~Chen, T.~Zhang, Y.~Hui, T.~Berg{-}Kirkpatrick, and S.~Dubnov, ``Large-scale contrastive language-audio pretraining with feature fusion and keyword-to-caption augmentation,'' in \emph{{IEEE} International Conference on Acoustics, Speech and Signal Processing {ICASSP} 2023, Rhodes Island, Greece, June 4-10, 2023}.\hskip 1em plus 0.5em minus 0.4em\relax {IEEE}, 2023, pp. 1--5. [Online]. Available: \url{https://doi.org/10.1109/ICASSP49357.2023.10095969}
\BIBentrySTDinterwordspacing

\bibitem{ssim}
\BIBentryALTinterwordspacing
Z.~Wang, A.~C. Bovik, H.~R. Sheikh, and E.~P. Simoncelli, ``Image quality assessment: From error visibility to structural similarity,'' \emph{{IEEE} Transactions on Image Processing}, vol.~13, no.~4, pp. 600--612, 2004. [Online]. Available: \url{https://doi.org/10.1109/TIP.2003.819861}
\BIBentrySTDinterwordspacing

\bibitem{si-sdr}
\BIBentryALTinterwordspacing
J.~L. Roux, S.~Wisdom, H.~Erdogan, and J.~R. Hershey, ``{SDR} - half-baked or well done?'' in \emph{{IEEE} International Conference on Acoustics, Speech and Signal Processing, {ICASSP} 2019, Brighton, United Kingdom, May 12-17, 2019}.\hskip 1em plus 0.5em minus 0.4em\relax {IEEE}, 2019, pp. 626--630. [Online]. Available: \url{https://doi.org/10.1109/ICASSP.2019.8683855}
\BIBentrySTDinterwordspacing

\bibitem{si-sdri}
\BIBentryALTinterwordspacing
Y.~Z. Isik, J.~L. Roux, Z.~Chen, S.~Watanabe, and J.~R. Hershey, ``Single-channel multi-speaker separation using deep clustering,'' in \emph{Interspeech 2016, 17th Annual Conference of the International Speech Communication Association, San Francisco, CA, USA, September 8-12, 2016}, N.~Morgan, Ed.\hskip 1em plus 0.5em minus 0.4em\relax {ISCA}, 2016, pp. 545--549. [Online]. Available: \url{https://doi.org/10.21437/Interspeech.2016-1176}
\BIBentrySTDinterwordspacing

\end{thebibliography}

%
%
%
%

\end{document}